\documentclass[twocolumn,english,prl,showpacs]{revtex4}
\usepackage[colorlinks=true,urlcolor=blue,citecolor=blue,linkcolor=blue]{hyperref}
\usepackage{ucs}
\usepackage[utf8x]{inputenc}
\usepackage{amssymb}
\usepackage{graphicx}
\usepackage{amsmath,color}
\usepackage{mathrsfs}
\usepackage{babel}	
\usepackage{bbold}

\newcommand{\tr}{\mathop{\text{tr}}\nolimits}

\newcommand{\ket}[1]{|{#1}\rangle}
\newcommand{\bra}[1]{\langle{#1}|}

\newcommand{\p}{^{\vphantom{\dagger}}}
\newcommand{\h}{^{\dagger}}

\newcommand{\Up}{{\uparrow}}
\newcommand{\Dn}{{\downarrow}}

\begin{document}

\title{Efficient continuous-time quantum Monte Carlo algorithm for fermionic lattice models}

\author{Mauro Iazzi}
\author{Matthias Troyer}

\affiliation{Theoretische Physik, ETH Zurich, 8093 Zurich, Switzerland}

\begin{abstract}
Efficient continuous time quantum Monte Carlo (CT-QMC) algorithms that do not suffer from time discretization errors  have become the state-of-the-art for most discrete quantum models. They have not been widely used yet for fermionic quantum lattice models, such as the Hubbard model, nor other fermionic lattice systems due to a suboptimal scaling of $O(\beta^3)$ with inverse temperature $\beta$, compared to the linear scaling of discrete time algorithms. Here we present a CT-QMC algorithms for fermionic lattice systems that matches the scaling of discrete-time methods but is more efficient and free of time discretization errors. This provides an efficient simulation scheme that is free from the systematic errors opening an avenue to more precise studies of large systems at low and zero temperature.
\end{abstract}

\maketitle

Monte Carlo simulations of quantum systems are often performed using an imaginary time path integral formulation \cite{PhysRev.91.1291} to map the partition function of the quantum system to that of an equivalent classical one \cite{Suzuki1977a,Barker:1979is}. These imaginary time paths, whose extent corresponds to the inverse temperature $\beta=1/k_BT$ are then sampled using Monte Carlo methods. Path integrals are usually formulated on a discrete imaginary time mesh with nonzero time step $\Delta_\tau$ in order to regularize the generally fractal paths. An extrapolation of the measured observables to $\Delta_\tau\rightarrow0$ is then required to obtain accurate results corresponding to those of the original quantum system.

For discrete quantum lattice models, some quantum Monte Carlo (QMC) algorithms exist that are free from time discretization errors, such as Handscomb's method for Heisenberg spin models \cite{PSP:2060252} or its generalization, the stochastic series expansion (SSE) algorithm \cite{Sandvik:1991ht}. They avoid an explicit introduction of time discretization by working with a Taylor expansion.  For these and other discrete quantum lattice models one can also avoid time discretization errors in a path-integral formulation by realizing that the lattice structure already provides a regularization of the path integral.

Over the last two decades a new category of path-integral quantum Monte Carlo algorithms has thus been developed that work directly in the continuous time limit $\Delta_\tau\rightarrow0$, removing the need for an extrapolation and often significantly speeding up the simulations. The first of these continuous-time quantum Monte Carlo (CT-QMC) algorithms have been for quantum spin systems and boson systems \cite{Beard1996,Prokofev:1998gz}. Combined with efficient non-local update algorithms, such as cluster updates \cite{Evertz:2003ch} or the worm algorithm \cite{Prokofev:1998gz} the gains in efficiency resulting from continuous time schemes are such that the simulation of unfrustrated spins and bosons is now considered a solved problem.

The generalization of CT-QMC to fermionic systems has been less straightforward, but has finally been achieved over the past decade by using time-dependent perturbation theory formulations of continuous time path integrals \cite{Gull2011}. The first fermionic CT-QMC algorithm for lattice models \cite{Rombouts:1999ip} was followed by a number of algorithms for fermionic quantum impurity problems \cite{PhysRevB.72.035122,Werner:2006ko,Gull:2008cm,Gull2011}. These algorithms have been widely employed as quantum impurity solvers~\cite{Gull2011} {\it i.e.} for simulating an open system embedded in a non-interacting bath. They have replaced discrete time algorithms as the state of the art method by being significantly more efficient, avoiding the need to extrapolate in $\Delta_\tau$, and allowing the simulation of a much wider class of models. In particular, they have revolutionized the solution of the quantum impurity problem arising from self-consistent dynamical mean field (DMFT) theories \cite{Metzner:1989bz,Georges:1992kt,Georges:1996hv}  and their cluster extensions \cite{Maier:2005et}. They allow the accurate simulation of much larger systems for Hubbard-type problems  \cite{Fuchs2011,PhysRevB.83.075122,PhysRevB.88.155108} and enable to go beyond density-density interactions by allowing the full Coulomb interaction to be included \cite{PhysRevB.74.155107} thus opening the way to realistic materials simulations by multi-orbital DMFT \cite{Kotliar2006}.

Despite their enormous success for quantum impurity models, CT-QMC methods are rarely used for fermionic lattice models~\cite{Fuchs2011}. There, discrete time methods \cite{Blankenbecler1981} are still the method of choice because of better scaling behavior. Existing fermionic CT-QMC algorithms all scale as $O(\beta^3V^3)$ with the inverse temperature $\beta$ and the lattice size $V$,  since these algorithms require operations to be performed on square matrices with dimension $O(\beta V)$.  For quantum impurity problems, which are described by time-dependent actions after integrating out the bath, also discrete time algorithms have the same scaling  \cite{Hirsch1986}. However, for quantum lattice models, discrete time algorithms exist that operate on $\beta/\Delta_\tau$ matrices of dimension $O(V)$ and the effort thus scales only as $O(\beta V^3)$. The substantially reduced scaling ensured a significant competitive advantage of the discrete time approach. 

Several (unpublished) attempts have been made to develop efficient CT-QMC methods for quantum lattice models. Na\"ive approaches have failed, giving either a worse sign problem or a $O(\beta V^4)$, erasing the advantage from the better scaling in temperature already for medium size systems. In this Letter we show how to overcome the issues and present a CT-QMC algorithm that has linear scaling in $\beta$ while retaining the cubic $O(V^3)$ complexity with respect to volume.

While our method is more general, we will -- for the sake of simplicity -- focus our presentation on the Hubbard model. The Hubbard model is the prototypical example of a strongly correlated fermionic system. It consists of spin-$\frac12$ fermionic particles that can hop between neighbouring sites of a lattice and repel via a contact interaction. The full Hamiltonian $H$ is given by a sum of the noninteracting and interacting parts $H=H_0+H_I$
\begin{eqnarray}
 H_0 &=& -\sum_{x,y, \sigma} t_{xy}c\h_{x\sigma}c\p_{y\sigma} \\ 
 H_I &=& U\sum_x \hat{h}_x \equiv U\sum_x \left(\hat{n}_{x\Up}-\frac12\right)\left(\hat{n}_{x\Dn}-\frac12\right).
\end{eqnarray}
Here the creation operators $c\h_{x\sigma}$ introduces a new fermion at site $x$ with spin $\sigma$ and the annihilation operator $c\p_{x\sigma}$ likewise removes one particle, subject to the canonical anti-commutation relations $\{c\h_{x\sigma},c\p_{y\sigma'}\}=\delta_{xy}\delta_{\sigma\sigma'}$. The occupation number for site $x$ is given by the density operator $\hat{n}_{x\sigma}=c\h_{x\sigma} c\p_{x\sigma}$. The tunneling matrix $t_{xy}=t_{yx}$ \footnote{Or more generally $t_{xy}=t_{yx}^*$ in the presence of a gauge field. Since  gauge fields generally introduce a phase problem we restrict ourselves to the case of real $t_{xy}$} gives the energy associated with the hopping from site $y$ to site $x$. When two particles of opposite spin are present in the same site, the they repel each other with energy $U$.

To perform the CT-QMC we perform a time-dependent perturbation expansion in  $H_I$ obtaining \cite{Prokofev1996,Gull2011}
\begin{eqnarray}\label{eq:ct}
 Z&=& \tr e^{-\beta H} = \tr\left[ e^{-\beta H_0} \mathcal{T} e^{-\int_0^\beta d\tau H_I(\tau) }\right] =  \\ &=&  \sum_{k=0}^{\infty} \frac{1}{k!}  \tr\left[  
\mathcal{T} e^{-\beta H_0} \int_0^\beta d\tau_1\ldots\int_0^\beta d\tau_{k} \prod_{i=1}^k(-H_I(\tau_i))\right] \nonumber 
\end{eqnarray}
where $H_I(\tau) = e^{\tau H_0}H_Ie^{-\tau H_0}$ is the perturbation term $H_I$ in the interaction representation.
%At this first stage, the expansion is a sum over all possible expansion orders $n$ and sets of interaction times $\tau_1, \ldots, \tau_n$. 
We then further expand the partition function as 
\begin{eqnarray}\label{eq:ctz}
Z = \sum_k \int_0^\beta d\tau_1\int_{\tau_1}^\beta\ldots\int_{\tau_{k-1}}^\beta d\tau_{k} \sum_{x_1,\ldots,x_k} w(c)
\end{eqnarray}
where the factor $\frac{1}{k!}$  is taken care of by time ordering $\tau_1 <\tau_2 < \ldots < \tau_k$.  $c=\{ (x_1,\tau_1),\ldots, (x_k\tau_k)\}$ denotes a continuous time path integral configuration with $k$ vertices and weight
\begin{equation}\label{eq:ctaux_weight}
 w(c) = \tr\left[e^{-\beta H_0} \prod_{i=1}^k (-U\hat{h}_{x_k}(\tau_k))\right] 
\end{equation}
CT-QMC now proceeds by sampling from all possible configurations $c$ according to their weight $w(c)$. 

The structure of the factors in Eq.~(\ref{eq:ctaux_weight}) allows the weight to be rewritten as~\cite{Koonin1997}
\begin{equation}
 w(c) = (-U)^{k}\det(1+\mathbf{B}(c,\beta)), %e^{-\beta \mathbf{H_0}}\prod_i \mathbf{h}(\tau_i, x_i))
 \label{eq:weight}
\end{equation}
where single particle propagator matrix $\mathbf{B}$ is given by
\begin{multline}
 \mathbf{B}(c,\beta) = e^{-\beta \mathbf{H}_0}\prod_i {\mathbf{h}}(\tau_i, x_i) =\\=
  e^{-(\beta-\tau_{k}) \mathbf{H}_0} {\mathbf{h}}(x_{k}) \ldots e^{-(\tau_2-\tau_1) \mathbf{H}_0} \mathbf{h}(x_1) e^{-\tau_1 \mathbf{H}_0}
\end{multline}
where $\mathbf{H}_0$ is a $2V\times 2V$ matrix with elements
$
\mathbf[{H}_0]_{x\sigma,y\sigma'}=t_{xy}\delta_{\sigma,\sigma'}
$
and the matrix ${\mathbf{h}}(x_i)$ is given by 
\begin{multline}
  [{\mathbf{h}}(x_i)]_{x\sigma,y\sigma'} = \delta_{xx_i}\delta_{x_iy}\delta_{\sigma\Up}\delta_{\sigma'\Up}+\\+
  \delta_{xx_i}\delta_{x_iy}\delta_{\sigma\Dn}\delta_{\sigma'\Dn}-\frac12\delta_{xy}\delta_{\sigma\sigma'},
\end{multline}
and ${\mathbf{h}}(\tau_i, x_i)=e^{\tau_i \mathbf{H}_0}{\mathbf{h}}(x_i)e^{-\tau_i \mathbf{H}_0}$ is its time-displaced counterpart. In the case of the Hubbard the matrix $\mathbf{B}$ decomposes into two block matrices for each spin species, giving
\begin{equation}
 w(c) = (-U)^{k}\det(1+\mathbf{B}^\Up(c,\beta))\det(1+\mathbf{B}^\Dn(c,\beta))
\end{equation}

The factor $(-U)^k$ introduces a sign problem for positive $U$, since any configuration with an odd number of vertices will have negative weight. On a bipartite lattice with nearest neighbour hoppings this trivial minus sign problem can be removed by mapping the repulsive model into an attractive one with interaction via a particle-hole transformation of the spin-down fermions $c\p_{x\Dn} \rightarrow (-1)^xc\h_{x\Dn}$. This transformation changes the sign of U and thus removes this trivial sign problem, while the sublattice dependent sign avoids changing the sign of the kinetic energy. However a sign problems can still appear from the determinant. 

This formulation of our algorithm is similar in spirit to the interaction-representation (CT-INT) algorithm for quantum impurity problems \cite{PhysRevB.72.035122,Gull2011} and can be considered a lattice CT-INT (or LCT-INT). Although not required, it has been found to be advantageous to instead use an auxilliary field decomposition~\cite{Hirsch1983} to remove the trivial sign for fermionic CT-QMC algorithms, leading the continuous time auxiliary field (CT-AUX) algorithm in the case of quantum impurity models \cite{Gull:2008cm,Gull2011}. The CT expansion is applied with $\hat{h}(x) = \left(\hat{n}_{x\Up}\hat{n}_{x\Dn}-1\right)$. In our algorithm we can introduce an auxiliary field $\rho$ giving an LCT-AUX representation
\begin{equation}\label{eq:aux}
 \left(1-\hat{n}_{x\Up}\hat{n}_{x\Dn}\right) = \frac{1}{2}\sum_{\rho=\pm 1} \left(1+\rho\hat{n}_{x\Up}\right)\left(1-\rho\hat{n}_{x\Dn}\right)
\end{equation}
so that every vertex now also carries a spin degree of freedom and configurations are of the form $\{(x_i, \tau_i, \rho_i)\}_{i=1\ldots k}$. 
%As the spin up and spin down parts of the action are no more equal, there is still a possibility for a non trivial sign problem to arise unless a symmetry explicitly prevents it as in the case of a half-filled Hubbard model after the particle-hole transformation. 
The auxiliary field dependent matrix  ${\mathbf{h}}(x_i, \rho_i)$ now becomes
\begin{multline}
 [{\mathbf{h}}(x_i, \rho_i)]_{x\sigma,y\sigma'} = \delta_{xy}\delta_{\sigma\sigma'}+\\+\rho_i\delta_{xx_i}\delta_{x_iy}\delta_{\sigma\Up}\delta_{\sigma'\Up} -\rho_i\delta_{xx_i}\delta_{x_iy}\delta_{\sigma\Dn}\delta_{\sigma'\Dn},
\end{multline}
and we end up with a weight similar to Eq. (\ref{eq:weight}), but using ${\mathbf{h}}(x_i, \rho_i)$ instead of ${\mathbf{h}}(x_i)$.
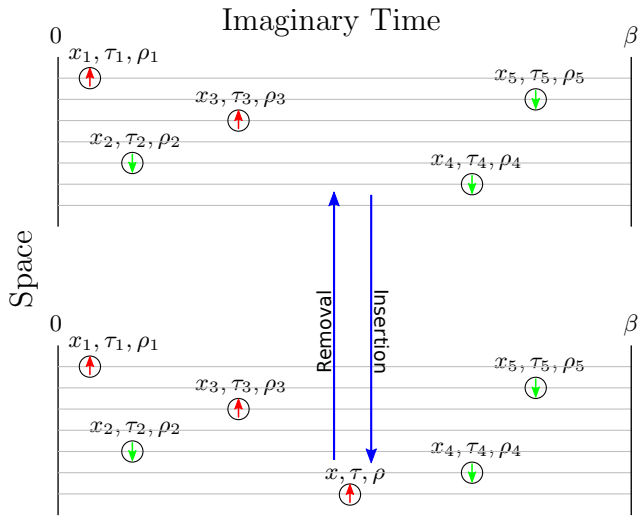
\begin{figure}
 \begin{picture}(240,190)
  \put(0,80){\rotatebox{90}{\large Space}}
  \put(80,185){\large Imaginary Time}
  %upper
  \put(15,180){$0$}
  \put(232,180){$\beta$}
  \put(22,174){$x_1,\tau_1,\rho_1$}
  \put(30,142){$x_2,\tau_2,\rho_2$}
  \put(70,158){$x_3,\tau_3,\rho_3$}
  \put(159,134){$x_4,\tau_4,\rho_4$}
  \put(183,166){$x_5,\tau_5,\rho_5$}
  %lower
  \put(15,71){$0$}
  \put(232,71){$\beta$}
  \put(22,65){$x_1,\tau_1,\rho_1$}
  \put(30,33){$x_2,\tau_2,\rho_2$}
  \put(70,49){$x_3,\tau_3,\rho_3$}
  \put(159,25){$x_4,\tau_4,\rho_4$}
  \put(183,57){$x_5,\tau_5,\rho_5$}
  \put(119,15){$x,\tau,\rho$}
  \put(15,0){\includegraphics[scale=0.5]{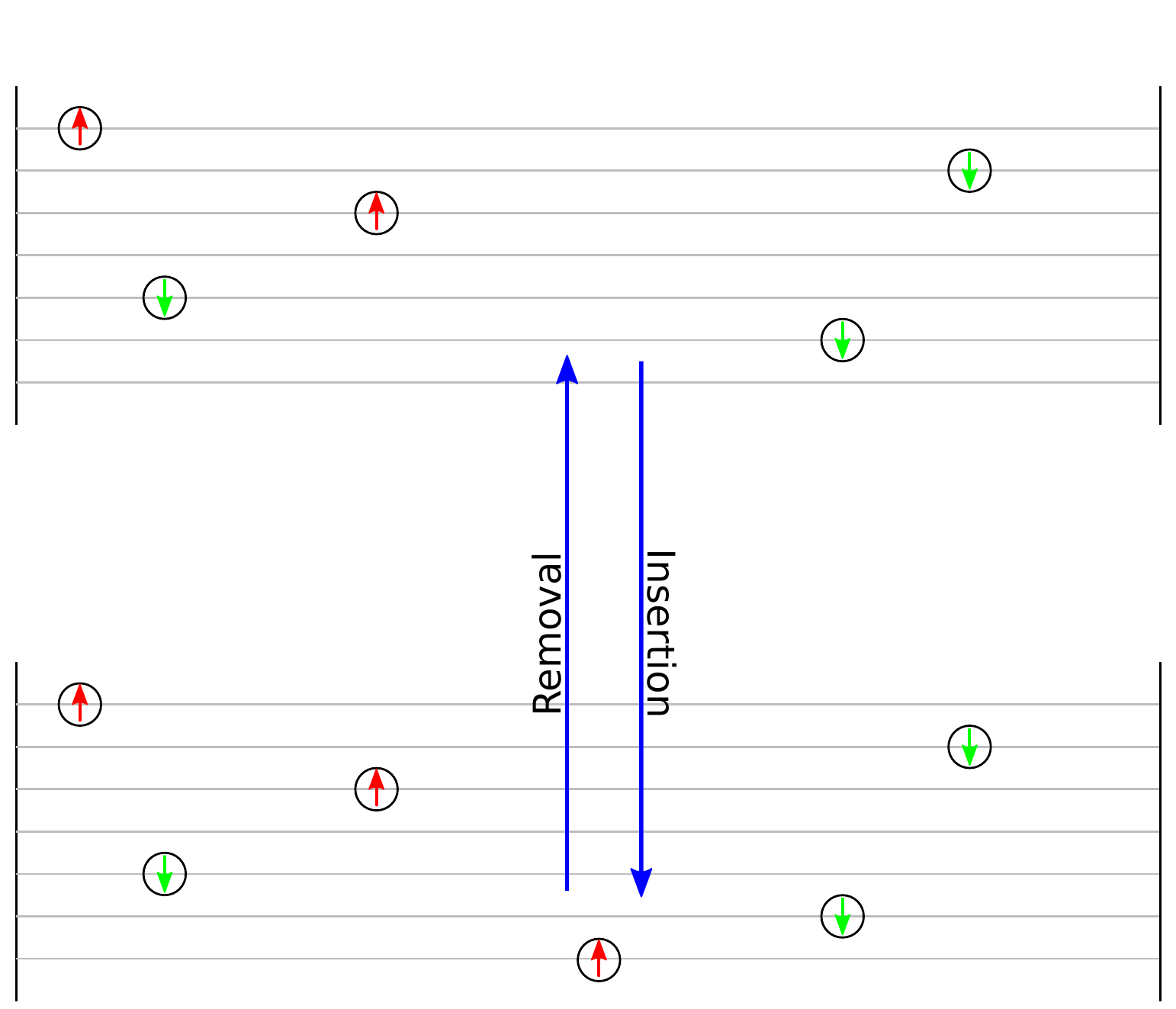}}
 \end{picture}
 \caption{Sketch of the insertion and removal updates. The $5$-vertex configuration $c$ can be modified adding a vertex at site $x$ and time $\tau$ with spin $\rho$. This leads to a proposed configuration $c'$. The reverse move consists of removing the vertex $(x, \tau, \rho)$. These two basic moves (insertion and removal) are sufficient to reach any term in the series \eqref{eq:ct} from any other.}\label{fig:updates}
\end{figure}

To ergodically sample all possible terms in the series \eqref{eq:ct} it is sufficient to implement two types of Monte Carlo updates: insertion and removal of a vertex. They change the order $k$ by $\pm1$ and are illustrated  in Fig. \ref{fig:updates}. Starting from a configuration $c$ with $k$ vertices one proposes to randomly insert a new vertex with auxiliary field $\rho$ at position $x$ and time $\tau$. The probability of accepting the new configuration $c'$ is given, using the Metropolis algorithm~\cite{Metropolis1953},  as $\min(1, R)$ with \footnote{The probability of inserting at a uniformly chosen time and location is $d\tau/(V\beta)$ while the probability of removing one of $k+1$ vertices is $1/(k+1)$. The infinitesimal $d\tau$ cancels with infinitesimals in the weights, the a-priori probability $1/2$ of choosing one of the two values of the auxiliary field cancels with a factor $1/2$ in Eq. (\ref{eq:aux}) and the factor $\beta V/(k+1)$ enters the acceptance ratio.}
\begin{equation}
 R = \frac{\beta V|U|}{k+1} \cdot \frac{\det\left[1+\mathbf{B}(c',\beta)\right]}{\det\left[1+\mathbf{B}(c,\beta)\right]}.
\end{equation}
Conversely, the probability of removing a vertex is $\min(1,1/R)$. The same acceptance ratio is derived for the LCT-INT version of the algorithm.

For any observable $\mathcal{O}$ one can write an estimator $O(c)$ which must be averaged to obtain an estimate of the quantum expectation value $\langle\mathcal{O}\rangle$. For equal-time observables such as densities, kinetic, and interaction energy, these are simple functions of the matrix $\mathbf{B}$. The single particle density matrix -- or equivalently equal time Green function -- $G(x,\sigma;y,\sigma') = \langle c\p_{x\sigma}c\h_{y\sigma'}\rangle$ is estimated by measuring the matrix elements $ \mathbf{G}_{x\sigma,y\sigma'}$ of the matrix
\begin{equation}
 \mathbf{G} = \frac{\mathbf{B}(c,\beta)}{1+\mathbf{B}(c,\beta)}.
\end{equation}
The kinetic energy estimator is simply $E_0 = \tr(\mathbf{H}_0 \mathbf{G})$, while the interaction energy is given by $E_I = U\sum_x \mathbf{G}_{x\Up,x\Up}\mathbf{G}_{x\Dn,x\Dn}$.

\begin{figure}
 \begin{picture}(230,170)
   \put(0,0){\includegraphics[width=\columnwidth]{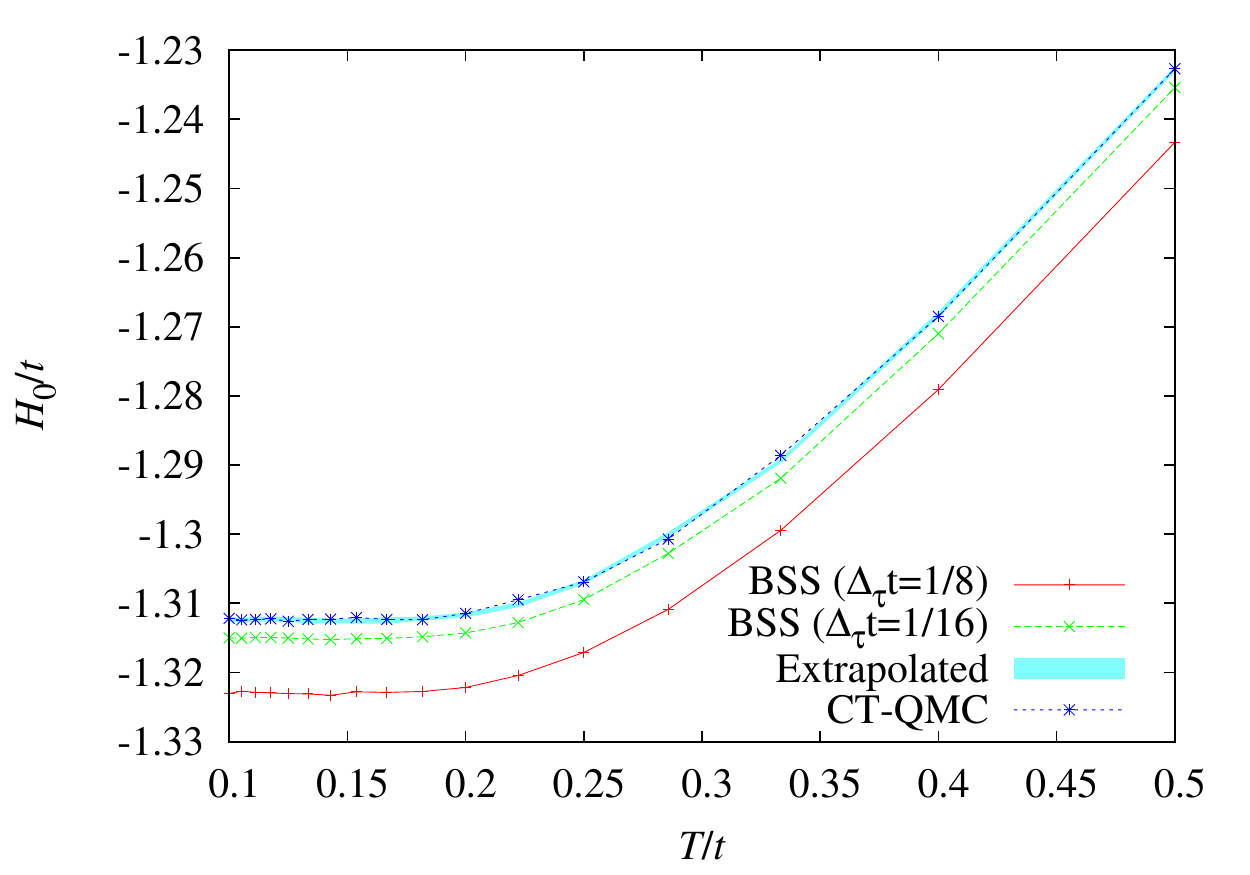}}
   \put(35,80){\includegraphics[width=0.45\columnwidth]{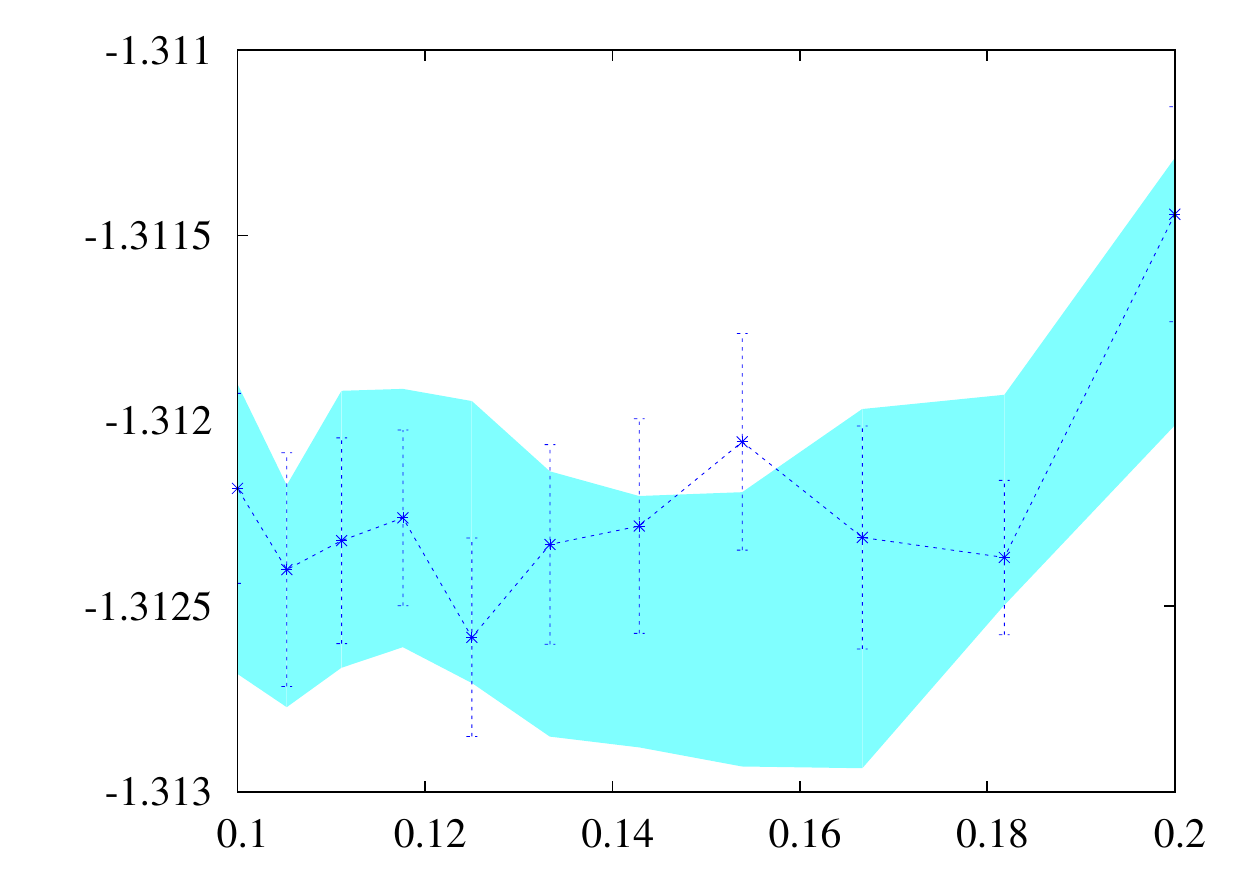}}
 \end{picture}
 \begin{picture}(230,170)
   \put(0,0){\includegraphics[width=\columnwidth]{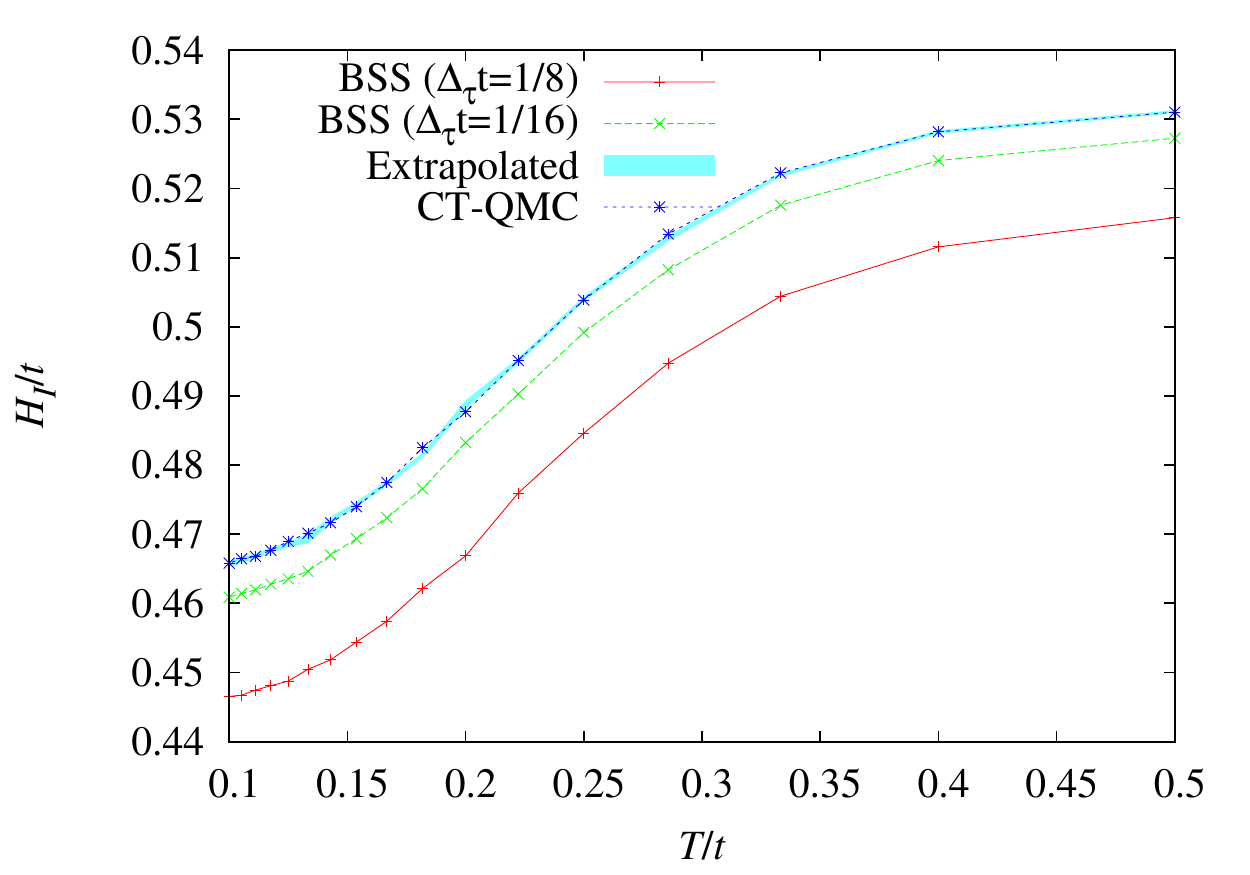}}
   \put(120,30){\includegraphics[width=0.45\columnwidth]{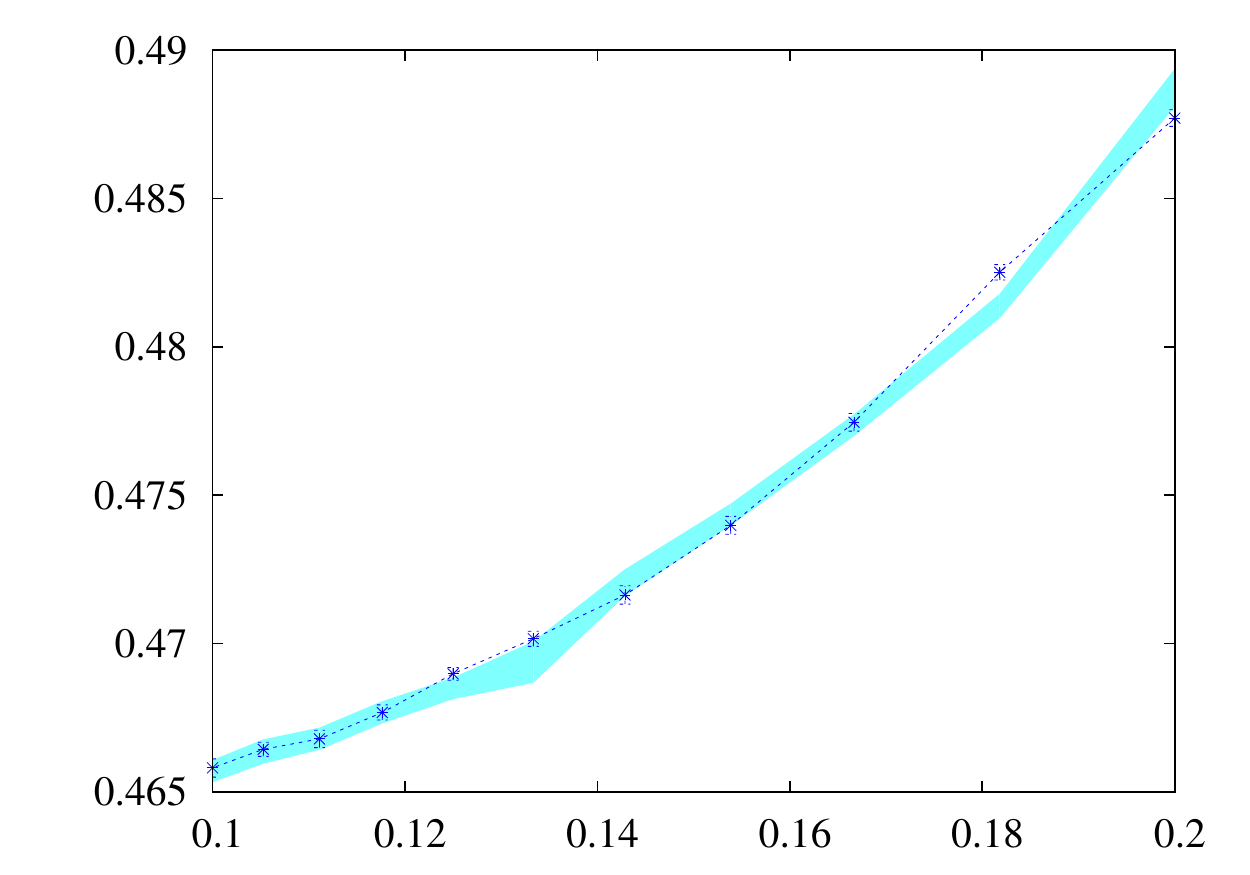}}
 \end{picture}
 \caption{Kinetic energy and interaction energy for a $4\times 4$ Hubbard plaquette at half filling with $U=4t$, computed using the discrete time BSS algorithms for various values of $\Delta_\tau t$ and LCT-AUX. The inset compares the CT-QMC results to BSS results extrapolated to $\Delta_\tau \rightarrow 0$.} \label{fig:results}
\end{figure}

To demonstrate the reliability and performance of our algorithm we compared it to the discrete time BSS algorithm~\cite{Blankenbecler1981}, which has so far been state of the art. Instead of starting from a continuous time representation (\ref{eq:ct}), this algorithm is based on the Suzuki Trotter formula
\begin{equation}
 e^{-\beta H} = \left(e^{-\Delta_\tau H_0}e^{-\Delta_\tau H_I}\right)^N + O(\beta \Delta_\tau^2).
\end{equation}
which entails a so-called Trotter error due to time discretization that is quadratic in the time step $\Delta_\tau=\beta/N$. Figure \ref{fig:results}  shows  that the results from our CT-QMC algorithm agree perfectly with those obtained by extrapolating the finite-$\Delta_\tau$ results obtained with the BSS algorithm to $\Delta_\tau\rightarrow0$. The  advantage of our algorithm is that it does not require this extrapolation in $\Delta_\tau$. This is particularly important for quantities such as the specific heat or double occupancy, which in the vicinity of phase transitions are very sensitive to the Trotter error. Equilibration and ergodicity features are also comparable, with autocorrelation times for the observables being very similar. This is easily understood as both the present method and the state-of-the-art BSS scheme only employ local updates (vertex insertion/removal and single spin flip respectively).

\begin{figure}
 \includegraphics[width=\columnwidth]{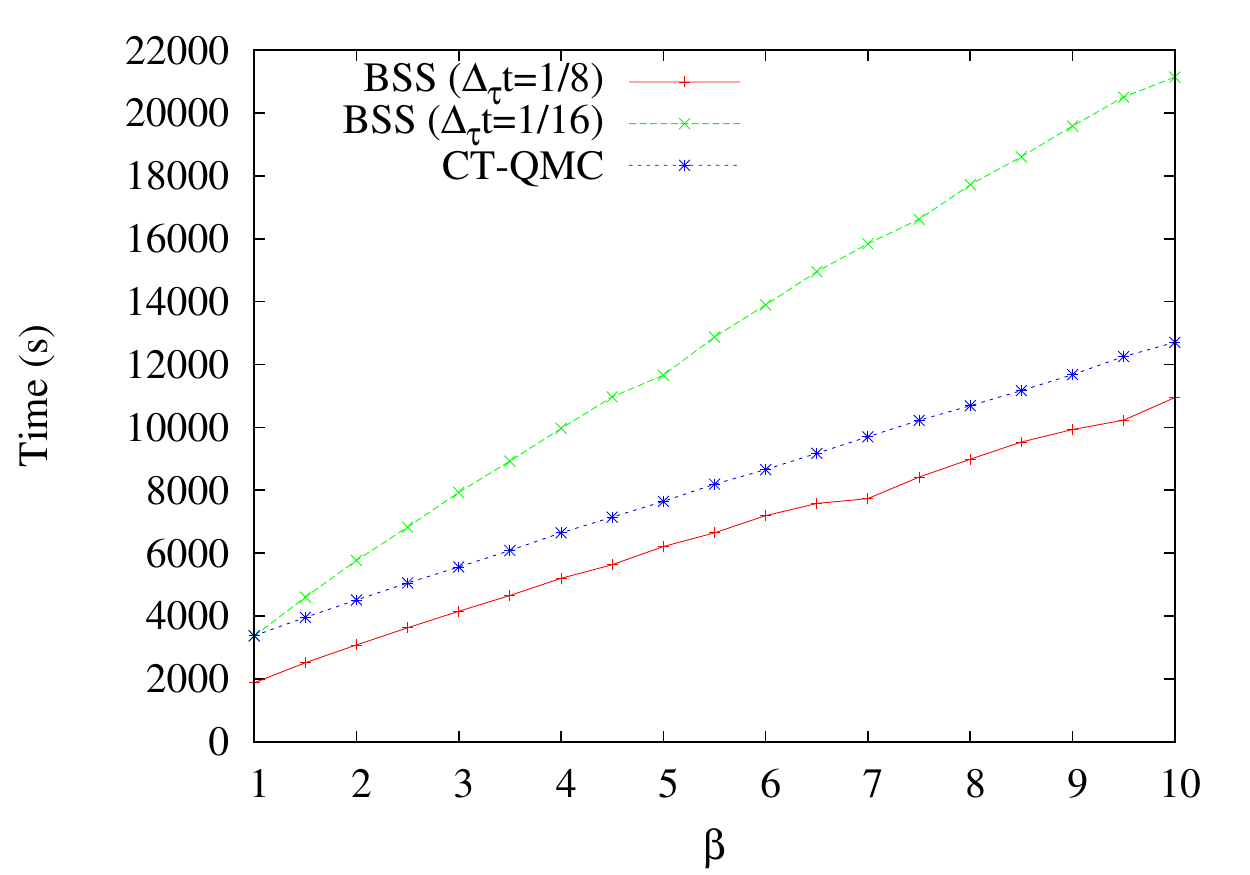}
 \caption{Computational time of the simulations from Fig.~\ref{fig:results} as a function of $\beta$. Both the present method and discrete time algorithm have been run for $10^6$ sweeps for thermalization then $10^6$ sweeps while measuring observables. Both codes have been similarly optimized with delayed rank-$1$ updates. The linear dependence is clear in both pictures. The slope of the CT scheme is the same as the DT with $\Delta_\tau t = 1/8$ because the average number of vertices in a time slice of size $t$ is around $120$, i.e. roughly the same as $8V$=128. The CT code extrapolates to a much larger constant for $\beta\rightarrow 0$ mostly due to several allocations per update which have not been optimized.}
 \label{fig:scaling}
\end{figure}

The main computational effort in the algorithm is calculating the matrix $\mathbf{B}(\beta)$ and  its changes when performing updates. Na\"ive multiplications of $k$ matrices of dimension $O(V)$ would result in an effort of $k O(V^3)$. Considering that the number of vertices $k$ grows with $\beta$, $U$  and $V$ we obtain a scaling of $O(\beta V^4)$, which is worse than the  of the discrete time BSS algorithm. To achieve an overall $O(\beta V^3)$ scaling  our algorithm works in the {\em eigenbasis} of $H_0$. Since the weight $w(c)$ is a determinant, it is unaffected by a basis change. Basis dependent quantities, such as observables, can be obtained via a rotation of the density matrix $\mathbf{G}$.

Diagonalizing  $\mathbf{H}_0 = \mathbf{U}\p \mathbf{E} \mathbf{U}\h$, where $\mathbf{E}$ is diagonal, the factors  $e^{-(\tau_{i+1}-\tau_i) \mathbf{H}}$ also become diagonal matrices $e^{-(\tau_{i+1}-\tau_i) \mathbf{E}}$. The other factors are of the form $\delta_{kk'} + \gamma \mathbf{U}\h_{kx_i}\mathbf{U}\p_{x_ik'}$, which is an identity matrix and an outer product of two vectors. Given this decomposition into sparse matrices and an outer product, the matrix multiplications can be performed with an effort $O(V^2)$, thus recovering the  $O(\beta V^3)$ scaling of the BSS algorithm. 

The product of matrices is in general an ill-conditioned matrix. To prevent numerical errors from creeping into the simulation, a {stabilization procedure} must be used, as explained in the Supplementary Material.

Common optimization techniques that are employed in other fermionic QMC algorithms can be applied here as well. Fast updates can be performed due to the fact that $\mathbf{B}(c)$ before the insertion (or removal) of a vertex, and $\mathbf{B}(c')$ after it, differ only by a single factor ${\mathbf{h}}(x, \tau)$, which is a diagonal matrix having all coefficients equal to $1$ except a single one~\cite{Blankenbecler1981}. Several updates can also be combined in a {delayed update} scheme \cite{Alvarez2008,PhysRevB.83.075122}. We implemented delayed updates for both our discrete time and continuous time codes.

Our performance measurements in Fig. \ref{fig:scaling}  confirm the  linear scaling in $\beta$. We have compared LCT-AUX with a BSS code using the same set of optimizations. In these conditions LCT-AUX performs as well as the BSS algorithm with a commonly used time step of $\Delta_\tau=1/8$.  Since the BSS simulations have to be repeated for several values of $\Delta_\tau$ and extrapolated, our unoptimized CT-QMC is already faster than a full discrete time calculation.

Using projections from a trial wave function our algorithm can be used for ground state simulations, similar to the discrete time algorithms~\cite{Sugiyama1986} (See Supplementary Material). It can also be used as a a quantum impurity solver and used for DMFT calculations \cite{Metzner:1989bz,Georges:1992kt,Georges:1996hv,Maier:2005et,Kotliar2006,Khatami2010} by adding $V_b$  non-interacting bath sites. The complexity of such an algorithm is $O(\beta U V(V+V_b)^2)$, which for low temperatures can be better than the $O(\beta^3U^3V^3)$ scaling of other CT-QMC algorithms \cite{PhysRevB.72.035122,Werner:2006ko,Gull:2008cm,Gull2011} for Hubbard-type models. The time-dependent Green functions $G(\tau; x, y) = \langle \mathcal{T}[c\p_x(\tau)c\h_y(0)]\rangle$ required for DMFT can be measured using partial propagators $\mathbf{B}(c,\tau)$:
\begin{equation}
 \mathbf{G}(\tau; \{x_i, \tau_i\}) = \frac{\mathbf{B}(c,\tau)}{1+\mathbf{B}(c,\beta)}
\end{equation}
with
\begin{equation}
 \mathbf{B}(c, \tau) = e^{-\tau \mathbf{H}_0}\prod_{\tau_i<\tau} \mathbf{h}(\tau_i, x_i).
\end{equation}
In general it is more stable to measure its Fourier transform, using non-uniform fast Fourier transformations~\cite{Greengard2004}.

The algorithm presented here is not specific to the Hubbard model. The only requirements are that the noninteracting Hamiltonian can be diagonalized once at the start of the algorithm to obtain the eigenvector matrix $\mathbf{U}$ and that the interacting Hamiltonian can be decomposed into exponentials of quadratic operators, i.e. $e^{\mathbf{A_{_xy}}c\h_xc\p_y}$ for some matrix $\mathbf{A}$. This is in general possible for any four-fermions interaction of the type $H_I=\sum_{kk'pp'}c\h_kc\p_{k'}c\h_pc\p_{p'}$ including the Coulomb interaction. As for the discrete time scheme, local interactions will retain the $O(V^3)$ scaling, but this might change in the general case (see Supplementary Material for more discussion).

In summary, we have presented a continuous time QMC algorithm for fermionic lattice models that has the same scaling as discrete time methods. This closes the last prominent gap in the portfolio of CT-QMC algorithms, which have otherwise become the state of the art for bosons, quantum spins, and fermionic impurity problems. The main advantage of our algorithm is the absence of any time discretization error. This eliminates the need to either guess a small enough time step $\Delta_\tau$ or extrapolate from multiple simulations at different $\Delta_\tau$ to $\Delta_\tau\rightarrow0$ and leads to shorter simulation times. Our algorithm can profit from the same numerical optimizations previously developed for other fermionic QMC algorithms \cite{Blankenbecler1981,Alvarez2008,PhysRevB.83.075122} and can be used for finite temperature simulations, ground state calculations and quantum chemistry simulations and as a quantum impurity solver.

We acknowledge discussions with F.F. Assad and Lei Wang. This work was supported by the ERC Advanced Grant SIMCOFE, Microsoft Research, and the Swiss National Science Foundation through the National Competence Centers in Research NCCR QSIT and MARVEL. MT acknowledges hospitality of the Aspen Center for Physics, supported by NSF grant \# 1066293.

\appendix

\section{\large Supplementary Material}

\section{Stabilization procedure}

At the core of our algorithm is the calculation of the matrix
\begin{multline}
 \mathbf{B}(c,\beta) = e^{-\beta \mathbf{H}_0}\prod_i \mathbf{h}(\tau_i, x_i) =\\=
  e^{-(\beta-\tau_{k}) \mathbf{H}_0} \mathbf{h}(x_{k}) \ldots e^{-(\tau_2-\tau_1) \mathbf{H}_0} \mathbf{h}(x_1) e^{-\tau_1 \mathbf{H}_0}
\end{multline}

As the ratio between the largest and lowest eigenvalue diverges information about the lowest eigenvalues and eigenstates is lost when the ratio between smallest and largest eigenvalues  become of the order of the roundoff. Calculations of the determinant of $G(\beta)$ then becomes inaccurate.

Numerical stabilization of the product of matrices with an acceptable accuracy is made possible by periodically decomposing the intermediate result using a rank-revealing decomposition such as a singular value decomposition (SVD) or pivoting QR. One first multiplies all the vertices within a certain imaginary-time interval $\tau_S = \frac{\beta}{M}$, then decompose them
\begin{equation}
e^{-\Delta\tau \mathbf{H}_0}\prod_{\tau_i<\tau_S} \mathbf{h}(\tau_i, x_i) \rightarrow \mathbf{U}_1 \mathbf{D}_1 \mathbf{V}_1^t
\end{equation}
The vertices up to $2\tau_S$ are then multiplied by $U_1D_1$ and decomposed again.
\begin{equation}
\left(e^{-2\tau_S \mathbf{H}_0}\prod_{\tau_S<\tau_i<2\tau_S} \mathbf{h}(\tau_i, x_i) \right) \mathbf{U}_1 \mathbf{D}_1 \rightarrow \mathbf{U}_2 \mathbf{D}_2 \mathbf{V}_2^t
\end{equation}
The procedure is repeated until the full product has been performed
\begin{equation}
\mathbf{B}(c,\beta) = \mathbf{U}_M \mathbf{D}_M \mathbf{V}_M^t\ldots \mathbf{V}_1^t.
\end{equation}
The number of intervals $M$ should be chosen so that the condition number of the partial products can be stored within machine precision. Additionally another SVD is performed whenever more than a certain number of vertices $m$ is multiplied consecutively within the same interval.

An additional issue is present when the partial product $\mathbf{B}(c, \tau)$ has a high degeneracy (as is the case when few vertices are present). In this situation the numerical errors in the decompositions artificially lift the degeneracy and introduce spurious components in $\mathbf{B}$. To avoid this problem one can start the product from a random matrix $\mathbf{R}$ and finish it multiplying by $\mathbf{R}^{-1}$ obtaining $\mathbf{R}^{-1}\mathbf{B}\mathbf{R}$, which can be used in all the formulas for $\mathbf{B}$ with minimal modifications.

\section{Fast updates}

When a vertex is inserted at $x', \tau'$, a single matrix $\mathbf{h}$ is inserted in the expression for $\mathbf{B}$. This enables the use of {rank}-$1$ updates, via the matrix determinant lemma 
\begin{equation}
 \frac{\det(A+uv^t)}{\det(A)} = 1+v^t(1+A)^{-1}u.
\end{equation}
where $A$ is a matrix and $u, v$ two vectors. In this case the matrix is $\mathbf{B}$ and the two vectors are the $x'$-th column of the eigenvector matrix $\mathbf{U}$ evolved once backwards and once forward in time
\begin{multline}
 u = e^{-(\beta-\tau_{k}) \mathbf{H}_0} \mathbf{h}(x_{k}) \ldots e^{-(\tau_{j+1}-\tau_j) \mathbf{H}_0} \\ \mathbf{h}(x_j) e^{-(\tau_j-\tau') \mathbf{H}_0} \mathbf{U}_{kx'}
\end{multline}
where $\tau_j$ is the first vertex after $\tau'$, and
\begin{multline}
 v^t = \mathbf{U}^t_{x'k}  e^{-(\tau'-\tau_{j-1}) \mathbf{H}_0} \\ \mathbf{h}(x_{j-1}) \ldots e^{-(\tau_2-\tau_1) \mathbf{H}_0} \mathbf{h}(x_1) e^{-\tau_1 \mathbf{H}_0}
\end{multline}
The weight ratio can then be computed as
\begin{equation}
 1+v^t\mathbf{G}\mathbf{B}^{-1}u
\end{equation}
where the vector $\mathbf{B}^{-1}u$ can be efficiently computed as
\begin{multline}
 \mathbf{B}^{-1}u = e^{\tau_1 \mathbf{H}_0} \mathbf{h}^{-1}(x_1) e^{(\tau_2-\tau_1) \mathbf{H}_0} \ldots \mathbf{h}^{-1}(x_{j-1})\\ e^{(\tau'-\tau_{j-1}) \mathbf{H}_0} \mathbf{U}_{kx'}
\end{multline}
if the newly inserted vertex is near the time origin. For this reason it can be advantageous to only allow vertex insertion (or removal) between the temporal origin and a maximum time $\tau_w$, and move the origin periodically, shifting all the vertex times by a constant time interval. The acceptance ratio becomes
\begin{equation}
 R = \frac{\tau_w V|U|}{m+1} (1+v^t\mathbf{G}\mathbf{B}^{-1}u)
\end{equation}
where $\tau_w$ is the size of the time window where the updates can happen and $m$ is the number of vertices within the same window. We can see that, if the vertex is added at the origin, the weight ratio reduces to the one for the discrete time case
\begin{equation}
 \frac{w(c')}{w(c)} = -U\mathbf{G}_{x'x'}
\end{equation}
This feature can be leveraged by always inserting or removing new vertices at the time origin, which is shifted at each update. The shifting procedure can be performed without a full recomputation of the matrix $\mathbf{G}$ using the formula
\begin{equation}
 \mathbf{G}(\tau) = \mathbf{B}(\tau)\mathbf{G}(0)\mathbf{B}^{-1}(\tau)
\end{equation}
which links the Green function at time $\tau$ to the Green function at time $0$. If the number of vertices comprised between $0$ and $\tau$ is $m$, the evolution matrices $\mathbf{B}(\tau)$ and its inverse can be applied with $O(mV^2)$ operations, so we need to limit the maximum time shift so that $m$ is on average constant. Moreover, since this procedure is not numerically stable, one still needs to periodically recompute $\mathbf{B}$ and $\mathbf{G}$ from scratch.

\section{Ground state method}

As for the BSS algorithm, the present CT-QMC can be adapted to a \textit{projector method} for ground state fermions. In such a scheme one wants to evaluate the expectation values
\begin{equation} \label{eq:projection}
 \langle O \rangle = \lim_{\theta\rightarrow\infty}\frac{\bra{\psi_T}e^{-\frac{\theta}{2} \hat{H}}\hat{O}e^{-\frac{\theta}{2} \hat{H}}\ket{\psi_T}}{\bra{\psi_T}e^{-\theta \hat{H}}\ket{\psi_T}}
\end{equation}
where $\ket{\psi_T}$ is a trial wavefunction and $\lim_{\theta\rightarrow\infty}e^{-\theta \hat{H}}$ is the projector on the ground state. If the trial wavefunction is not orthogonal to the ground state Eq.~\eqref{eq:projection} gives the correct results for the expectation value of $O$. The trial wavefunction is chosen to be a Slater determinant of $p$ fermions described by the $p\times n$ matrix $\mathbf{P}$
\begin{equation}
 \ket{\psi_T} = \prod_p (\mathbf{P}_{p1}c\h_1 + \ldots + \mathbf{P}_{pn}c\h_n)\ket{0}
\end{equation}

It is easy to recognize that $\theta$ has the same function as the inverse temperature $\beta$ in the finite temperature scheme, so one can  rewrite the expansion \eqref{eq:ct} as
\begin{eqnarray}\label{eq:gct}
 Z&=& \bra{\psi_T}e^{-\theta H}\ket{\psi_T} = \bra{\psi_T} e^{-\theta H_0} \mathcal{T} e^{-\int_0^\theta d\tau H_I(\tau) }\ket{\psi_T} = \nonumber \\ &=&  \sum_{k=0}^{\infty} \frac{(-1)^k}{k!}  \bra{\psi_T} 
 e^{-\theta H_0} \mathcal{T}\int_0^\theta d\tau_1\ldots\int_0^\theta d\tau_{k} \prod_{i=1}^kH_I(\tau_i)\ket{\psi_T} = \nonumber \\
 &=& \sum_k \int_0^\theta d\tau_1\int_{\tau_1}^\theta\ldots\int_{\tau_{k-1}}^\theta d\tau_{k} \sum_{x_1,\ldots,x_k} w(c)
\end{eqnarray}
where the weight is now given only by the expectation value of the operator $B(c, \theta)$ over the trial state $\ket{\psi_T}$. Since the trial state is a Slater determinant, the weight can still be expressed as a determinant of a single particle matrix
\begin{equation}
 w(c) = \det(\mathbf{P}\p\mathbf{B}(c, \theta)\mathbf{P}\h).
\end{equation}
This weight can be computed using the same techniques s in the finite temperature case, however there are additional simplifications that can be exploited in the ground state formalism. If one splits the  evolution in two parts, from $0$ to $\tau$ and from $\tau$ to $\theta$, we can define
\begin{eqnarray}
 \mathbf{L}(\tau) &=& \mathbf{P} \mathbf{B}(c; \theta, \tau) \\
 \mathbf{R}(\tau) &=& \mathbf{B}(c; \tau, 0) \mathbf{P}\h
\end{eqnarray}
Performing a SVD of both $\mathbf{L}$ and $\mathbf{R}$ yelds
\begin{eqnarray}
 \mathbf{L} &=& \mathbf{V_L} \mathbf{D_L} \mathbf{U_L}\h \\
 \mathbf{R} &=& \mathbf{U_R} \mathbf{D_R} \mathbf{V_R}\h
\end{eqnarray}
where $\mathbf{V_{L/R}}$ are unitary $p\times p$ matrices, $\mathbf{D_{L/R}}$ are diagonal with $p$ elements and $\mathbf{U_{L/R}}$ are $n\times p$ matrices whose columns are orthonormal vectors. The equal-time Green function at time $\tau$ is then given by
\begin{multline}
 \mathbf{G}(\tau) = 1 - \mathbf{R}(\tau) [\mathbf{L}(\tau)\mathbf{R}(\tau)]^{-1} \mathbf{L}(\tau) =\\=
 1 - \mathbf{U_R} [\mathbf{U_L}\h\mathbf{U_R}]^{-1} \mathbf{U_L}\h
\end{multline}
Since one needs only $\mathbf{U_{L/R}}$ to compute the Green function, the stabilization procedure is considerably simplified, since one can discard the intermediate $\mathbf{D}$s and $\mathbf{V}$s. The fast-update method can still be implemented in this formalism by noting that
\begin{equation}
 \frac{\det(\mathbf{L}(\tau)(1+uv^t)\mathbf{R}(\tau))}{\det(\mathbf{L}(\tau)\mathbf{R}(\tau))} = 1 + v^t(1-\mathbf{G}(\tau))u
\end{equation}
which again only requires the knowledge of the $\mathbf{U}$s to be computed.

\section{Implicit and explicit bath}

For Dynamical Mean Field Theory (DMFT) and Dynamical Cluster Approximation (DCA) calculations, it is necesary to simulate a system where the noninteracting particle Green function includes a self energy $\mathbf{\Sigma}$
\begin{equation}
 \mathbf{G}(i\omega) = \frac{1}{i\omega - \mathbf{H}_0 + \mathbf{\Sigma}(i\omega)}
\end{equation}
where the self energy is calculated from previous iterations of the algorithm (see Ref.~\cite{} for a review of DMFT/DCA). This is the so-called \textit{implicit bath} scheme, as the effect of a thermal bath is simulated indirectly. Since the self-energy depends in general on the imaginary frequency, the noninteracting systems is not described by a Hamiltonian and thus the present method cannot be employed. Instead one can use CT-AUX and CT-INT or the Hisrch-Fye QMC (which is of discrete-time type), which have all the same scaling properties.

Another option to simulate a system with a given self-energy is to introduce a number $N_b$ of ancillary sites to the model, with creation and annihilation operators $a\h_{y\sigma}$ and $a\p_{y\sigma}$. The dynamics of these bath sites and their interaction with the target system are described by the auxiliary Hamiltonian
\begin{equation}
 \hat{H}_b = \sum_y^{N_b}\sum_\sigma \epsilon_{y\sigma}a\h_{y\sigma}a\p_{y\sigma} + \sum_{x\sigma,y\sigma'}V_{x\sigma,y\sigma'} (c\h_{x\sigma}a\p_{y\sigma'} + a\h_{y\sigma'}c\p_{x\sigma})
\end{equation}
The values of the bath energy levels $\epsilon_y$ and hopping parameters $V_{xy}$ are chosen so that the self energy induced by the bath into the system
\begin{equation}
 \mathbf{\Sigma}_b(i\omega) = \sum_{xx'y} \frac{V_{xy}V_{x'y}}{i\omega-\epsilon_y}
\end{equation}
fits the desired self-energy $\mathbf{\Sigma}$. The fit can be performed minimizing the Green function misfit
\begin{multline}
 \chi^2(\{\epsilon, V\}) = \sum_{i\omega} w(i\omega) \left|\frac{1}{i\omega-\mathbf{H}_0-\mathbf{\Sigma}(i\omega)} \right.+\\- \left. \frac{1}{i\omega-\mathbf{H}_0-\mathbf{\Sigma}_b(i\omega)}\right|^2
\end{multline}
with respect to the bath parameters. The weighting factors $w(i\omega)$ can be used to prioritize accuracy of the fit over relevant (usually lower) frequencies. Such \textit{explicit bath} scheme is indeed applicable to the present method.

\section{Local and non-local interactions}

The present method can be applied to both local and non-local interactions. For a generic density-density interaction of the form
\begin{equation}
 H_I = \sum_x \sum_y U(|x-y|) \left(\hat{n}_x-\frac12\right)\left(\hat{n}_y-\frac12\right)
\end{equation}
we obtain several types of vertices coresponding to all possible pairs of sites $(x, y)$ and the corresponding single-particle matrix $\mathbf{h}(x_i, y_i)$ will depend on both site indices. In contrast to the simple Hubbard model, each vertex has a different coefficient $U(|x_i-y_i|)$. Moreover in this case there will be vertices corresponding to interactions between same-spin particles. In such cases the matrix $\mathbf{h}(x_i, y_i)$ only acts on one component of the spin (e.g. only the up component) as a \textit{rank-$2$} matrix of the form
\begin{multline}
  [{\mathbf{h}}(x_i, y_i)]_{x\sigma,y\sigma'} = \delta_{xx_i}\delta_{x_iy}\delta_{\sigma\Up}\delta_{\sigma'\Up}+\\+\delta_{xy_i}\delta_{y_iy}\delta_{\sigma\Up}\delta_{\sigma'\Up}-\frac12\delta_{xy}\delta_{\sigma\sigma'},
\end{multline}
i.e. a diagonal matrix with all coefficients equal to $-\frac12$ except for the ones at $x_i$ and $y_i$ which have the sign flipped. In the energy eigenbasis this matrix becomes
\begin{multline}
  [{\mathbf{h}}(x_i, y_i)]_{p\sigma,q\sigma'} = \mathbf{U}\h_{px_i}\mathbf{U}\p_{x_iq}\delta_{\sigma\Up}\delta_{\sigma'\Up}+\\+\mathbf{U}\h_{py_i}\mathbf{U}\p_{y_iq}\delta_{\sigma\Up}\delta_{\sigma'\Up}-\frac12\delta_{pq}\delta_{\sigma\sigma'}
\end{multline}

The acceptance ratio for insertion of such a vertex becomes
\begin{equation}
 R = \frac{\beta N_p|U|}{k+1} \cdot \frac{\det\left[1+\mathbf{B}(c',\beta)\right]}{\det\left[1+\mathbf{B}(c,\beta)\right]}.
\end{equation}
where $N_p$ is the number of possible pairs that can be inserted. In the case of nearest neighbour interactions on a square lattice $N_p=8V$, with $2V$ coming all possible links and a factor $4$ for all possible spin combinations. For a generic all-to-all interaction $N_p=4V*(V-1) + V$.

The number of vertices in the simulation is proportional to the ground state interaction energy $\langle H_I \rangle$. As such it will still grow linearly for any local and quasi-local interaction, as well as with power law interactions decaying faster than $\frac{1}{V}$, resulting again in a scaling of $O(\beta U V^3)$ for the simulation. For Coulomb potentials the interaction energy can grow faster than $V$ i.e. be super-extensive and results in a worse scaling with volume.

\bibliographystyle{ieeetr}
\bibliography{biblio}

\end{document}